\documentclass[journal,draftcls,onecolumn,12pt,twoside]{IEEEtran}

\normalsize

\usepackage{cite}
\usepackage{amsmath,amssymb,amsfonts}
\usepackage{algorithmic}
\usepackage{subcaption}
\usepackage{graphicx}
\usepackage{textcomp}
\usepackage{xcolor}
\usepackage{float}
\usepackage{amsthm}
\usepackage{graphicx}
\usepackage{epstopdf}
\usepackage{amsmath,bm}
\usepackage{amsfonts}
\usepackage{amssymb}
\usepackage{color}
\usepackage{multirow}
\usepackage{multicol}
\usepackage{soul,xcolor}
\usepackage{overpic}
\definecolor{orange}{RGB}{0,112,192}

\newtheorem{theorem}{Theorem}

\newtheorem{example}{Example}

\newcommand{\vect}[1]{\mathbf{#1}}

\def\diag{\mathrm{diag}}

\def\Htran{\mbox{\tiny $\mathrm{H}$}}

\def\CN{\mathcal{N}_{\mathbb{C}}} %Complex Gaussian
\def\diag{\mathrm{diag}}
\begin{document}
\makeatletter
\newcommand*{\rom}[1]{\expandafter\@slowromancap\romannumeral #1@}
\makeatother

\title{The Bussgang Decomposition of Non-Linear Systems: Basic Theory and MIMO Extensions}

\author{\"Ozlem Tu\u{g}fe Demir,~\IEEEmembership{Member,~IEEE,} and
	Emil Bj\"ornson,~\IEEEmembership{Senior Member,~IEEE}\thanks{The authors are with the Department of Electrical Engineering
		(ISY), Link\"oping University, 581 83 Link\"oping, Sweden (e-mail: ozlem.tugfe.demir@liu.se, emil.bjornson@liu.se). The authors were partially supported by ELLIIT and the Wallenberg AI, Autonomous Systems and Software Program (WASP) funded by the Knut and Alice Wallenberg Foundation.}}

\maketitle

\vspace{-14mm}

Many of the systems that appear in various signal processing applications are non-linear, for example, due to hardware impairments such as non-linear amplifiers and finite-resolution quantization. The Bussgang decomposition is a popular tool for analyzing the performance of systems that involve such non-linear components. In a nutshell, the decomposition provides an exact probabilistic relationship between the output and the input of a non-linearity: the output is equal to a scaled version of the input plus uncorrelated distortion. The decomposition can either be used to compute exact performance results or lower bounds where the uncorrelated distortion is treated as independent noise. This lecture note explains the basic theory, provides key examples, extends the theory to complex-valued vector signals, and clarifies some potential misconceptions.

\section{Relevance}
\vspace{-2mm}
The origin of the decomposition is a technical report by Julian J. Bussgang from 1952 \cite{Bussgang1952a}. Interestingly, the decomposition is not explicitly stated in his report, but rather a consequence of his results.
In fact, it is mainly non-trivial extensions of his results that are utilized in current research; for example, applications to complex-valued multiple-input multiple-output (MIMO) systems are popular in the communication community.
There is no standard reference that presents and proves those extended results, and it can be hard to differentiate between which results are exact and which are mere approximations. This lecture note fills these gaps.

\section{Prerequisites}

This lecture note requires basic knowledge of random variables, linear algebra, signals and systems, and estimation theory.

\section{Original Bussgang Decomposition for Real Gaussian Random Variables}
In the original paper \cite{Bussgang1952a}, Bussgang considers two jointly Gaussian stationary random processes $f(t)$ and $g(t)$. The process $f(t)$ undergoes a non-linear memoryless distortion represented by the function $U(\cdot)$.
The resulting non-Gaussian random process is  
\begin{align}
F(t)=U\big(f(t)\big).	
\end{align}
Bussgang computed the cross-correlation of the two random variables obtained by sampling $F(t)$ and $g(t)$ at specific time instances. Let $x = f(t_1) \in \mathbb{R}$ and $y = g(t_2) \in \mathbb{R}$ denote the zero-mean Gaussian random variables obtained by sampling at time $t_1$ and $t_2$, respectively. 
Moreover, let $z=F(t_1)=U(x) \in \mathbb{R}$ be the sampled output of the non-linear distortion function.  We then have the following main result from \cite[Sec.~III]{Bussgang1952a}.
 	\vspace{-0.4cm}
\begin{theorem}[The Bussgang theorem] \label{theorem_bussgang_original}
The cross-correlation of $z=U(x)$ and $y$ is
\begin{align}
C_{zy}=\mathbb{E}\left\{U(x)y\right\}=\underbrace{\frac{\mathbb{E}\left\{U(x)x\right\}}{\mathbb{E}\left\{x^2\right\}}}_{\triangleq B}\mathbb{E}\left\{xy\right\}=BC_{xy},
\end{align}
where $B$ is called the Bussgang gain and $C_{xy}\triangleq\mathbb{E}\left\{xy\right\}$ is the cross-correlation of $x$ and $y$.
\end{theorem}

The Bussgang theorem shows that the cross-correlation between two Gaussian signals is the same before and after one of them has passed through a non-linear function, except for a scaling factor $B$. The value of $B$ depends on the choice of $U(\cdot)$ but the theorem holds for any function.
 
A consequence of Theorem~\ref{theorem_bussgang_original} for $y=x$ is that the output signal can be decomposed as 
\begin{equation} \label{eq:basic-decomposition}
z=U(x)=Bx+\eta,
\end{equation}
where $\eta$ is a zero-mean random variable that is uncorrelated to both $x$ and $y$.
This is the \emph{Bussgang decomposition} in its elementary form and shows that the output contains the useful part $Bx$ and the distortion part $\eta$. In other words, the output of a non-linear function is equal to a scaled version of the input plus the uncorrelated distortion $\eta$. Note that  $\eta$ and $x$ are not independent. Since $\eta = U(x)-Bx$ is a deterministic function of $x$, the distortion term is non-Gaussian distributed and statistically dependent on $x$. Even if the Bussgang decomposition is named after Bussgang, the result is not explicitly stated in \cite{Bussgang1952a}.

\vspace{-4mm}
\section{Bussgang Decomposition for Complex Random Variables }
\vspace{-2mm}
The Bussgang theorem was extended to the complex case in \cite{Minkoff1985a}.
We will present this result and then provide a direct proof that is inspired by \cite{Bjornson2019b}. For notational convenience, in the remainder of this lecture note, we use $C_{x}\triangleq\mathbb{E}\{ |x|^2\}$ to denote the power of a signal $x$ and we use $C_{xy}\triangleq\mathbb{E}\left\{xy^*\right\}$ to denote the cross-correlation between $x$ and $y$.
\vspace{-2mm}
\begin{theorem}[The complex Bussgang theorem] \label{th:complex-Bussgang}
Consider the jointly circularly symmetric complex Gaussian random variables $x\in \mathbb{C}$ and $y\in \mathbb{C}$. Let $z=U(x) \in \mathbb{C}$ be the output of a deterministic function. The cross-correlations $C_{zy}\triangleq\mathbb{E}\left\{zy^*\right\}$ and  $C_{xy}\triangleq\mathbb{E}\left\{xy^*\right\}$ are then related as
\begin{align}
C_{zy}=\mathbb{E}\left\{U\left(x\right)y^*\right\}=\underbrace{\frac{\mathbb{E}\left\{U(x)x^*\right\}}{\mathbb{E}\left\{|x|^2\right\}}}_{\triangleq B = C_{zx}/C_{x}}\mathbb{E}\left\{xy^*\right\}=BC_{xy}. \label{bussgang_complex}
\end{align}
\end{theorem}

\begin{IEEEproof}
We begin by decomposing $y$ into two parts:
\begin{align}
y=\frac{\mathbb{E}\left\{yx^*\right\}}{{\mathbb{E}\left\{|x|^2\right\}}}x+
\underbrace{\left( y - \frac{\mathbb{E}\{yx^*\}}{{\mathbb{E}\left\{|x|^2\right\}}}x \right)}_{\triangleq \epsilon}.  \label{emil_proof1}
\end{align}
Interestingly, this is equivalent to computing a minimum-mean squared error (MMSE) estimate of $y$ given $x$, with $\epsilon$ representing the estimation error.
Hence, it follows that the second part, $\epsilon$, in \eqref{emil_proof1} is uncorrelated with $x$:
\begin{align}
\mathbb{E}\left\{\epsilon x^*\right\}=\mathbb{E}\left\{\left(y-\frac{\mathbb{E}\left\{yx^*\right\}}{{\mathbb{E}\left\{|x|^2\right\}}}x\right)x^*\right\} = 
\mathbb{E}\{yx^*\} - 
\frac{\mathbb{E}\left\{yx^*\right\}}{{\mathbb{E}\left\{|x|^2\right\}}} \mathbb{E}\left\{|x|^2\right\}
=0.
\end{align}
Since $x$ and $y$ are jointly Gaussian, the fact that $x$ and $\epsilon$ are uncorrelated implies that they are also independent complex Gaussian variables.
By using the decomposition in \eqref{emil_proof1}, it follows that 
\begin{equation}
C_{zy}= \mathbb{E}\left\{U\left(x\right)y^*\right\} = 
\frac{\mathbb{E}\left\{U(x)x^*\right\}}{\mathbb{E}\left\{|x|^2\right\}}\mathbb{E}\left\{xy^*\right\} + \underbrace{\mathbb{E}\left\{U\left(x\right)\epsilon^*\right\}}_{=0} = BC_{xy}
\end{equation}
by using that the independence between $x$ and $\epsilon$ implies $\mathbb{E}\{U(x)\epsilon^*\}=\mathbb{E}\{U(x)\} \mathbb{E}\{\epsilon^*\}= 0$.
\end{IEEEproof}

The complex Bussgang theorem is the natural complex-valued extension of Theorem~\ref{theorem_bussgang_original}. The corresponding complex Bussgang decomposition is given by \eqref{eq:basic-decomposition} with the only exception that the Bussgang gain is now computed as $B= \frac{C_{zx}}{C_{x}}= \frac{\mathbb{E}\left\{U(x)x^*\right\}}{\mathbb{E}\left\{|x|^2\right\}}$ instead.

A first use case of the Bussgang decomposition is to quantify the signal-to-distortion ratio (SDR) at the output of the distortion function. The SDR is simply the power ratio of the desired signal $Bx$ to the additive distortion $\eta$:
\begin{align}
\mathrm{SDR}=\frac{\mathbb{E}\left\{|Bx|^2\right\}}{\mathbb{E}\left\{|\eta|^2\right\}}=\frac{|B|^2 C_{x}}{C_{z} -|B|^2 C_{x}},
\end{align}
where we have used that the additive distortion $\eta$ is uncorrelated with the desired signal $x$.

\begin{figure}
	\begin{center}
			\includegraphics[trim={0cm 0cm 1.3cm 0.6cm},clip,width=10cm]{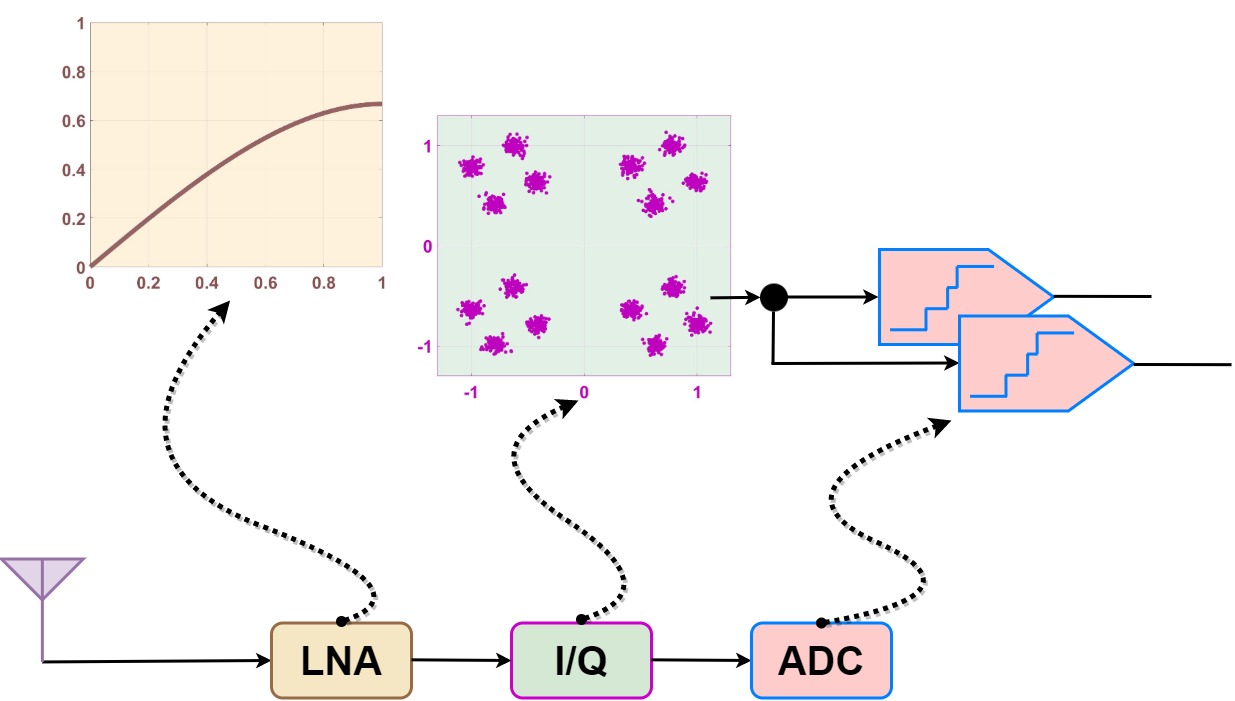}
	\end{center}
	\vspace{-0.5cm}
	\caption{Some common sources of hardware impairments in a wireless receiver.}\label{fig:1new}
		\vspace{-0.9cm}
\end{figure}

A second use case is to analyze the performance of a communication system where $x \sim \CN(0,C_{x})$ is the transmitted information signal. Suppose the received signal is noisy distorted signal $U(x)+w$, where $U(\cdot)$ models the hardware distortion and $w$ is thermal noise with power $\sigma^2$. 
The hardware distortion might, for example, be caused of a sequence of non-ideal blocks in the receiver hardware \cite{Schenk2008a}, as illustrated in Fig.~\ref{fig:1new}. The first block is the low-noise amplifier (LNA) that can distort both the amplitude and phase of the input signal. In the yellow figure, the amplitude distortion is exemplified and clipping occurs for input signals with large amplitudes. The second block is the in-phase/quadrature (I/Q) demodulator that might have mismatches between its branches leading to I/Q imbalance. In the green curve, the effect of I/Q imbalance is shown on a QPSK constellation where the actual transmitted points are affected by leakage from the mirror subcarriers.
Finally, in the analog-to-digital converter (ADC) block, the real and imaginary parts of the received signal are quantized to be represented by a finite number of bits. Quantization distortion is inevitable even if a large number of ADC bits are used \cite{Fletcher2007a,Bai2013a}.
We can use the Bussgang decomposition in \eqref{eq:basic-decomposition} to rewrite the received signal as 
\begin{equation}
U(x)+w = \underbrace{Bx}_{\textrm{Desired signal}}+\underbrace{\eta + w}_{\textrm{Uncorrelated signal}}.
\end{equation}
This signal contains a desired part $Bx$ and an uncorrelated additive ``noise'' term $\eta + w$. Since the latter term is uncorrelated with $x$, we can utilize the \emph{Worst case uncorrelated additive noise} theorem from \cite{Hassibi2003a} to compute an achievable data rate. That theorem says that the worst distribution of $\eta + w$ from a rate perspective is independent complex Gaussian, in which case the rate is
\begin{equation} \label{eq:Bussgang-rate}
\log_2 \left( 1+ \frac{\mathbb{E}\left\{|Bx|^2\right\}}{\mathbb{E}\left\{|\eta|^2\right\}+\sigma^2} \right)= \log_2 \left( 1+ \frac{|B|^2 C_{x}}{C_{z} -|B|^2 C_{x} + \sigma^2} \right) \quad \textrm{bit per channel use}.
\end{equation}
One can possibly achieve a larger rate than \eqref{eq:Bussgang-rate}, by somehow making use of the information content in $\eta$. But  we achieve \eqref{eq:Bussgang-rate} if we treat  $\eta$ as independent Gaussian noise in the decoder.

\vspace{-2mm}
\subsection{Alternative Computation of the Bussgang Gain and Two Examples}
\vspace{-1mm}
If the distortion function $U(x)$ is differentiable, there is an alternative way of computing the Bussgang gain that might be easier. We will exemplify how to compute it in the real-valued case where $x \sim \mathcal{N}(0,C_x)$ has the probability density function (PDF) $p(x) = \frac{1}{\sqrt{2\pi C_{x}}} e^{-x^2/(2C_{x})}$. Note that the derivative of this PDF is $p'(x) = - \frac{x}{C_{x}} p(x)$. We can then rewrite the Bussgang gain as
\begin{equation} \label{eq:alternative-computation}
B = \frac{\mathbb{E}\{ U(x) x \}}{C_{x}} =\int_{-\infty}^{\infty}    \frac{U(x) x}{C_{x}} p(x) dx  \overset{(a)}{=}   - \int_{-\infty}^{\infty} U(x) p'(x) dx  \overset{(b)}{=}  \int_{-\infty}^{\infty} U^{\prime}(x) p(x) dx = \mathbb{E} \{ U^{\prime}(x) \},
\end{equation}
where we identify $p'(x)$ in $(a)$ and integrate by parts to get $(b)$.
The last expression in \eqref{eq:alternative-computation} reveals that the Bussgang gain can be also computed as the expected value of the first derivative of the distortion function. This result is a special case of Price's Theorem \cite[Example 9-17]{papoulis2002}.

\vspace{-2mm}

\begin{example}[One-bit quantization] Consider a real-valued signal $x\sim\mathcal{N}(0,C_{x})$ that enters the non-linear distortion function $U(x)=\mathrm{sgn}(x)$, which represents one-bit quantization. The Bussgang gain can then be found as $B=\mathbb{E}\{U^{\prime}(x)\}=2\mathbb{E}\{\delta(x)\}=2p(0)= \sqrt{\frac{2}{\pi C_{x}}}$, where $\delta(x)$ is the Dirac function.
The same Bussgang gain can be computed as $B=\frac{\mathbb{E}\left\{U(x)x\right\}}{\mathbb{E}\left\{x^2\right\}}=\frac{\mathbb{E}\left\{|x|\right\}}{C_{x}}$.
\end{example}

\vspace{-2mm}

A similar alternative way of computing the Bussgang gain exists in the complex-valued case, where the derivative of the distortion function $U(x)$ is defined as \cite{McGee1969}:
\begin{align} \label{eq:complex-derivative}
& \frac{\partial U(x)}{\partial x}=\frac{1}{2}\left(\frac{\partial U(x)}{\partial \Re\left\{x\right\}}-j\frac{\partial U(x)}{\partial \Im\left\{x\right\}}\right).
\end{align}
One can then show that the Bussgang gain can be computed as  \cite{McGee1969}
\begin{equation}
B = \mathbb{E}\left\{ \frac{\partial U(x)}{ \partial x}\right\}.
\end{equation}

\vspace{-3mm}

\begin{example}[Third-order non-linearity] Consider a complex-valued signal $x\sim\CN(0,C_{x})$ that enters the third-order non-linear distortion function $U(x)=|x|^2x$, which might model a non-linear amplifier \cite{Bjornson2019b, Demir2020}. The Bussgang gain can be obtained as $B=\frac{\mathbb{E}\left\{|x|^4\right\}}{\mathbb{E}\left\{|x|^2\right\}}=2 C_{x}$. The same number is found by evaluating $B=\mathbb{E}\left\{ \frac{\partial U(x)}{ \partial x}\right\} = \mathbb{E}\{ 2|x|^2 \} = 2C_{x}$ using \eqref{eq:complex-derivative}.
\end{example}

\vspace{-2mm}

\vspace{-5mm}
\subsection{Additive Quantization Noise Model is Nothing But Bussgang Decomposition}
\vspace{-1mm}

The Bussgang decomposition is unique in the sense that it is the only decomposition $U(x)=Bx+\eta$ of a distorted signal having the property that the additive distortion noise $\eta$ is uncorrelated with the desired signal $x$. No other value of $B$ can be used to achieve that.

One seemingly different decomposition is the Additive Quantization Noise Model (AQNM) originally proposed in \cite{Fletcher2007a} to model quantization errors.
This model is sometimes described as an alternative decomposition, however, AQNM is nothing but the Bussgang decomposition for quantization. In \cite[Lemma 1]{Fletcher2007a}, a scalar quantizer function $\mathcal{Q}(\cdot)$ is considered, which has the property $\mathbb{E}\left\{x|\mathcal{Q}(x)\right\}=\mathcal{Q}(x)$, which means that each quantization interval is represented by its mean value.
When the input is $x\sim \CN(0,C_{x})$, it is shown that the output can be expressed as a summation of a scaled version of $x$ plus an uncorrelated distortion term $\eta$ as follows: \vspace{-2mm}
\begin{align}
z=\mathcal{Q}(x)=(1-\beta)x+\eta, \label{aqnm}
\end{align} 
where $\beta=\mathbb{E}\left\{\left|x-z\right|^2\right\}/C_{x}$ and $\mathbb{E}\left\{\left|\eta\right|^2\right\}=\beta(1-\beta)C_{x}$. 

We will show that \eqref{aqnm} equals the Bussgang decomposition $x=Bx+\eta$, where the Bussgang gain $B=C_{zx}/C_{x}$ equals $1-\beta$. Using the assumption $\mathbb{E}\left\{x|\mathcal{Q}(x)\right\}=\mathcal{Q}(x)$ from \cite{Fletcher2007a}, we have
\begin{align}
C_{zx}=\mathbb{E}\left\{\mathcal{Q}(x)x^*\right\}=\mathbb{E}\left\{\mathbb{E}\left\{\mathcal{Q}(x)x^*|\mathcal{Q}(x)\right\}\right\}=\mathbb{E}\left\{\mathcal{Q}(x)\mathcal{Q}^*(x)\right\}=C_{z}.
\end{align} 
By utilizing this result, the scaling  $1-\beta$ in \eqref{aqnm} can be rewritten as
\begin{align}
1-\beta=1 - \frac{\mathbb{E}\left\{\left|x-z\right|^2\right\}}{C_{x}} = 1-\frac{C_{x}+C_{z}-C_{zx}-C_{zx}^*}{C_{x}}=\frac{C_{zx}}{C_{x}}=B.
\end{align}
Hence, the AQNM is a special case of the Bussgang decomposition for distortion functions that satisfy a particular condition. The bottomline is that the Bussgang decomposition is unique but the value of $B$ depends on the distortion function.

\section{Extension to MIMO Systems}

In recent years, it has become popular to analyze MIMO systems that are subject to hardware impairments, in particular, in MIMO communications \cite{Bjornson2014a,Xu2019,Bai2013a}. In this part, we extend the Bussgang results to be applicable to such cases.

Consider two jointly circularly symmetric Gaussian random vectors ${\bf x}\sim\CN\left({\bf 0}, {\bf C}_{x}\right)$ and ${\bf y}\sim\CN\left({\bf 0}, {\bf C}_{y}\right)$, which both have length $M$. The correlation matrices are denoted as ${\bf C}_{x}=\mathbb{E}\{ \vect{x} \vect{x}^{\Htran} \}$ and ${\bf C}_{y}=\mathbb{E}\{ \vect{y} \vect{y}^{\Htran} \}$ and are assumed to have full rank. The cross-correlation matrix is denoted as ${\bf C}_{xy}=\mathbb{E}\left\{{\bf x}{\bf y}^{\Htran}\right\}$.  Using this notation, we can generalize the Bussgang theorem as follows.
 	\vspace{-0.2cm}
\begin{theorem}[Bussgang Theorem for MIMO Distortions] \label{th:Bussgang-MIMO}
Consider the jointly circularly symmetric Gaussian random vectors ${\bf x}$ and ${\bf y}$. Let ${\bf U}:\mathbb{C}^{M} \rightarrow\mathbb{C}^{M}$ denote a distortion function and ${\bf z}={\bf U}({\bf x})$ is the distorted signal when using $\vect{x}$ as input.
The cross-correlation matrix ${\bf C}_{zy}=\mathbb{E}\big\{{\bf z}{\bf y}^{\Htran}\big\}$ of ${\bf z}$ and ${\bf y}$ is a linear transformation of the cross-correlation matrix ${\bf C}_{xy}$ of $\vect{x}$ and $\vect{y}$:
\begin{align}
{\bf C}_{zy}= {\bf C}_{zx}{\bf C}_{x}^{-1} {\bf C}_{xy}. \label{bussgang_multi}
\end{align}
\end{theorem}
 	\vspace{-0.4cm}
\begin{IEEEproof}
The proof is a matrix extension of the proof of Theorem~\ref{th:complex-Bussgang}.
Let us express ${\bf y}$ as a summation of the MMSE estimate of it given ${\bf x}$ and the estimation error $\bm{\epsilon} \in \mathbb{C}^{M}$:
\begin{align}
{\bf y}={\bf C}_{yx}{\bf C}_{x}^{-1}{\bf x}+\bm{\epsilon}, \label{multi_proof}
\end{align}
where $\bm{\epsilon}$ is defined as $\bm{\epsilon} = {\bf y} - {\bf C}_{yx}{\bf C}_{x}^{-1}{\bf x}$. If we multiply both sides of \eqref{multi_proof} by ${\bf x}^{\Htran}$ from the right and take the expectation, we obtain
\begin{align}
{\bf C}_{yx}={\bf C}_{yx}{\bf C}_{x}^{-1}{\bf C}_{x}+\mathbb{E}\big\{\bm{\epsilon}{\bf x}^{\Htran}\big\}={\bf C}_{yx}+\mathbb{E}\big\{\bm{\epsilon}{\bf x}^{\Htran}\big\}, \label{multi_proof2}
\end{align}
from which it follows that $\mathbb{E}\big\{\bm{\epsilon}{\bf x}^{\Htran}\big\}={\bf 0}$. Hence, $\bm{\epsilon}$ and ${\bf x}$ are uncorrelated, which implies that they are also independent since these are jointly Gaussian variables.
Finally, we obtain \eqref{bussgang_multi} as
\begin{equation}
{\bf C}_{zy} = \mathbb{E} \{ \vect{z} \vect{y}^{\Htran} \} = 
\mathbb{E} \{ \vect{z} \vect{x}^{\Htran} \} {\bf C}_{x}^{-1} {\bf C}_{yx}^{\Htran}
+ \mathbb{E} \{ \vect{z} \bm{\epsilon}^{\Htran} \} = {\bf C}_{zx}{\bf C}_{x}^{-1} {\bf C}_{xy}
\end{equation}
by utilizing that ${\bf C}_{yx}^{\Htran} = {\bf C}_{xy}$ and that $\mathbb{E}\big\{{\bf z}\bm{\epsilon}^{\Htran}\big\}={\bf 0}$ since $\vect{z}$ and $\bm{\epsilon}$ are independent.
\end{IEEEproof}

From this theorem we notice that the Bussgang gain is represented by the matrix
\begin{equation} \label{eq:Bussgang-matrix}
{\bf B}={\bf C}_{zx}{\bf C}_{x}^{-1}
\end{equation}
and we call it a MIMO extension since the distortion function takes multiple inputs and provide multiple outputs. It is possible to extend the result to case where ${\bf C}_{x}$ is rank-deficient, in which case the inverse in \eqref{eq:Bussgang-matrix} is replaced by a pseudo-inverse; see  \cite[Section \rom{2}.A]{Bjornson2019b} for details.

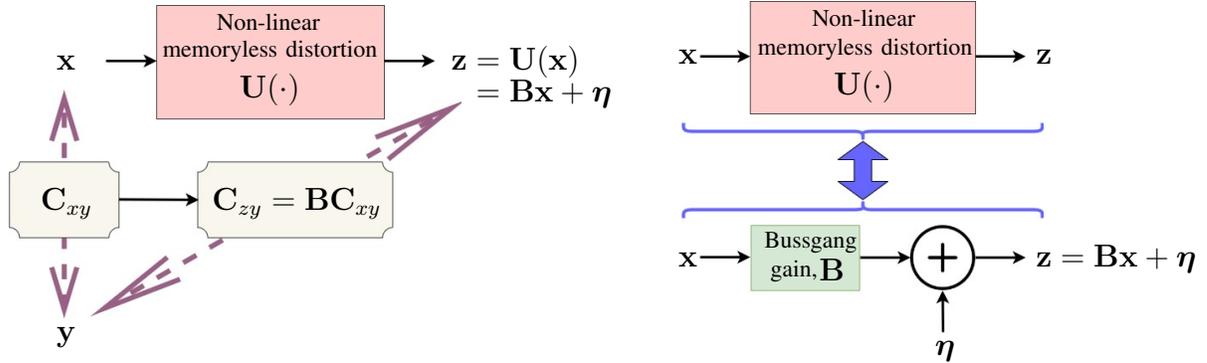
\begin{figure}
 	\begin{subfigure}{0.47\textwidth}
 		\begin{center}
 			\begin{overpic}[width=6cm,tics=10]{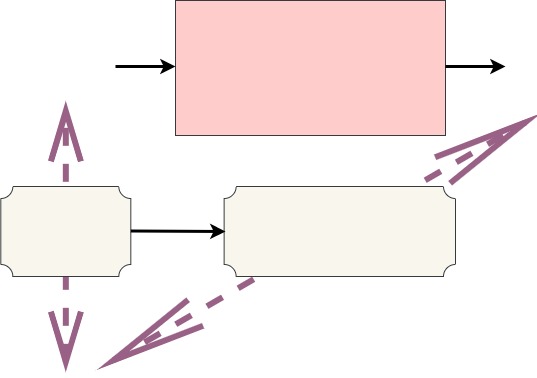}
 				\put(10.5,56){${\bf x}$}
 				\put(10.5,-4){${\bf y}$}
 				\put(46,65){\footnotesize Non-linear}
 				\put(34,59){\footnotesize memoryless distortion}
 				\put(51,50){${\bf U}(\cdot)$}
 				\put(98,56){${\bf z}={\bf U}({\bf x})$}
 				\put(103.5,49){$={\bf B}{\bf x}+\bm{\eta}$}
 				\put(7,25){${\bf C}_{xy}$}
 				\put(45,25){${\bf C}_{zy}={\bf B}{\bf C}_{xy}$}
 			\end{overpic}
 		\end{center}
 		\caption{Bussgang decomposition for jointly Gaussian random vectors ${\bf x}$ and ${\bf y}$.} \label{fig:1a}
 	\end{subfigure}
 	\hspace*{\fill} % separation between the subfigures
 	\begin{subfigure}{0.51\textwidth}
 		\begin{center}
 			\begin{overpic}[width=4.8cm,tics=10]{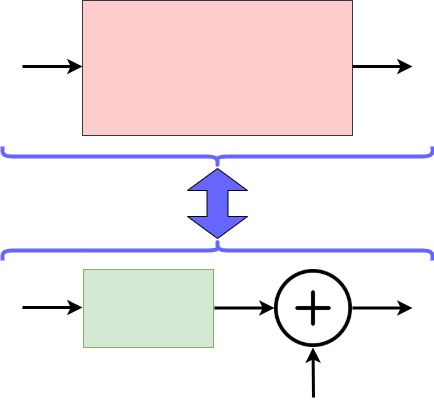}
 				\put(-1,76.5){${\bf x}$}
 				\put(-1,20.5){${\bf x}$}
 				\put(98,76.5){${\bf z}$}
 				\put(98,20.5){${\bf z}={\bf B}{\bf x}+\bm{\eta}$}
 				\put(70,-4.2){$\bm{\eta}$}
 				\put(22.8,25.5){\footnotesize Bussgang}
 				\put(24.5,17.5){\footnotesize gain,}
 				\put(38, 16.5){${\bf B}$}
 				\put(36,87){\footnotesize Non-linear}
 				\put(21,79){\footnotesize memoryless distortion}
 				\put(42,68){${\bf U}(\cdot)$}
 			\end{overpic} \vspace{-2mm}
 		\end{center}
 		\caption{Generalized Bussgang decomposition for non-Gaussian random vector ${\bf x}$.}\label{fig:1b}
 	\end{subfigure}
 \vspace{-0.4cm}
 	\caption{Bussgang decomposition for non-linear memoryless distortion function ${\bf U}(\cdot)$.} \label{fig:1}
 	 \vspace{-0.8cm}
 \end{figure}

A consequence of Theorem~\ref{th:Bussgang-MIMO} is the Bussgang decomposition for MIMO functions: 
\begin{equation} \label{eq:Bussgang-decomp-MIMO}
{\bf z}={\bf U}({\bf x}) = {\bf B}{\bf x}+\bm{\eta},
\end{equation}
where the additive distortion term $\bm{\eta}$ is uncorrelated both 
with ${\bf x}$ and any other random vector ${\bf y}$ that is correlated with ${\bf x}$.
This result is illustrated in Fig.~\ref{fig:1}(a).

\subsection{Element-Wise Distortion for MIMO Systems}

The Bussgang decomposition for MIMO functions has been widely used to model the hardware impairments in multiple-antenna communication systems \cite{Bai2013a,Bjornson2014a}. In this case, $M$ is the number of receive antennas and the distortion function represents impairments in the antenna branches.
A common assumption is that there is no crosstalk between the branches, so that each one can be separately modeled in the way shown in Fig.~\ref{fig:1new}.
The distortion function then has the form
\begin{align} \label{eq:elementwise-distortion}
{\bf z}={\bf U}({\bf x})=\begin{bmatrix} U_1(x_1) \\[-2mm] \vdots \\[-2mm] U_M(x_M) \end{bmatrix},
\end{align}
where $x_m$ denotes the $m^{\textrm{th}}$ element of ${\bf x}$. Hence, each output is a distorted version of only the input having the same index.
We can then simplify the Bussgang matrix by utilizing Theorem~\ref{th:Bussgang-MIMO}. More precisely, it follows that ${\bf C}_{zx}={\bf D}{\bf C}_{x}$, where ${\bf D}=\diag(d_1,\ldots,d_M)$ is a diagonal matrix and $d_m %=\frac{C_{zx}(m,m)}{C_{x}(m,m)}
=\frac{\mathbb{E}\{ U_m(x_m) x_m^*\}}{\mathbb{E}\{|x_m|^2\}}$ is the Bussgang gain corresponding to the $m^{\textrm{th}}$ component of the distortion function, i.e., $z_m=U_m(x_m)$. Hence, the Bussgang gain matrix of the overall MIMO distortion becomes ${\bf B}={\bf C}_{zx}{\bf C}_{x}^{-1} = {\bf D}$ and we obtain the simplified Bussgang decomposition  
\begin{align} 
{\bf z}={\bf D}{\bf x}+\bm{\eta} = \begin{bmatrix} d_1 x_1 \\[-2mm] \vdots \\[-2mm] d_M x_M \end{bmatrix} + \bm{\eta}.
\end{align}
Hence, when an element-wise distortion function affects the Gaussian signal ${\bf x}$, the output ${\bf z}$ is an element-wise scaled version of ${\bf x}$ plus a distortion vector  $\bm{\eta}$ that is uncorrelated with ${\bf x}$.

\vspace{-2mm}

\subsection{Are the Elements of Distortion $\bm{\eta}$ Uncorrelated?}
\label{sec:are-uncorrelated}

Since the Bussgang gain matrix is diagonal when having element-wise distortions, one may tend to think that the elements of the distortion $\bm{\eta}$ will also be uncorrelated, so that we effectively get one separate Bussgang decomposition per received signal. 
However, this is generally not the case as we will show next. 
Let ${\bf C}_{\eta}=\mathbb{E}\{\bm{\eta}\bm{\eta}^{\Htran}\} \in \mathbb{C}^{M\times M}$ denote the correlation matrix of the distortion vector $\bm{\eta}$. Using the fact that $\bm{\eta}$ is uncorrelated with $\vect{x}$, it can be computed as
\begin{align}
{\bf C}_{\eta}={\bf C}_{z}-{\bf B}{\bf C}_{x}{\bf B}^{\Htran}.
\end{align}
Whenever the input signal $\vect{x}$ contains correlated elements, such that ${\bf C}_{x}$ is non-diagonal, the correlation matrix will likely also be non-diagonal. This is intuitively quite clear: If two (almost) identical signals are sent through identical hardware components, then the distortion should also be (almost) identical. This type of correlation typically appears in wireless communications since each receive antenna observes a different linear combination of the same transmitted information signals.
Some conditions for when the correlation can be neglected, so that ${\bf C}_{\eta}$ is approximately diagonal, are derived in \cite{Bjornson2019b}. However, it is rather common that the correlation is neglected without motivation (cf.~\cite{Bai2013a,Xu2019}), which might lead to substantial approximation errors.

\begin{figure}[t!]
	\vspace{-0.5cm}
	\begin{center}
		\includegraphics[trim={0.2cm 0.1cm 0.5cm 0.5cm},clip,width=9.2cm]{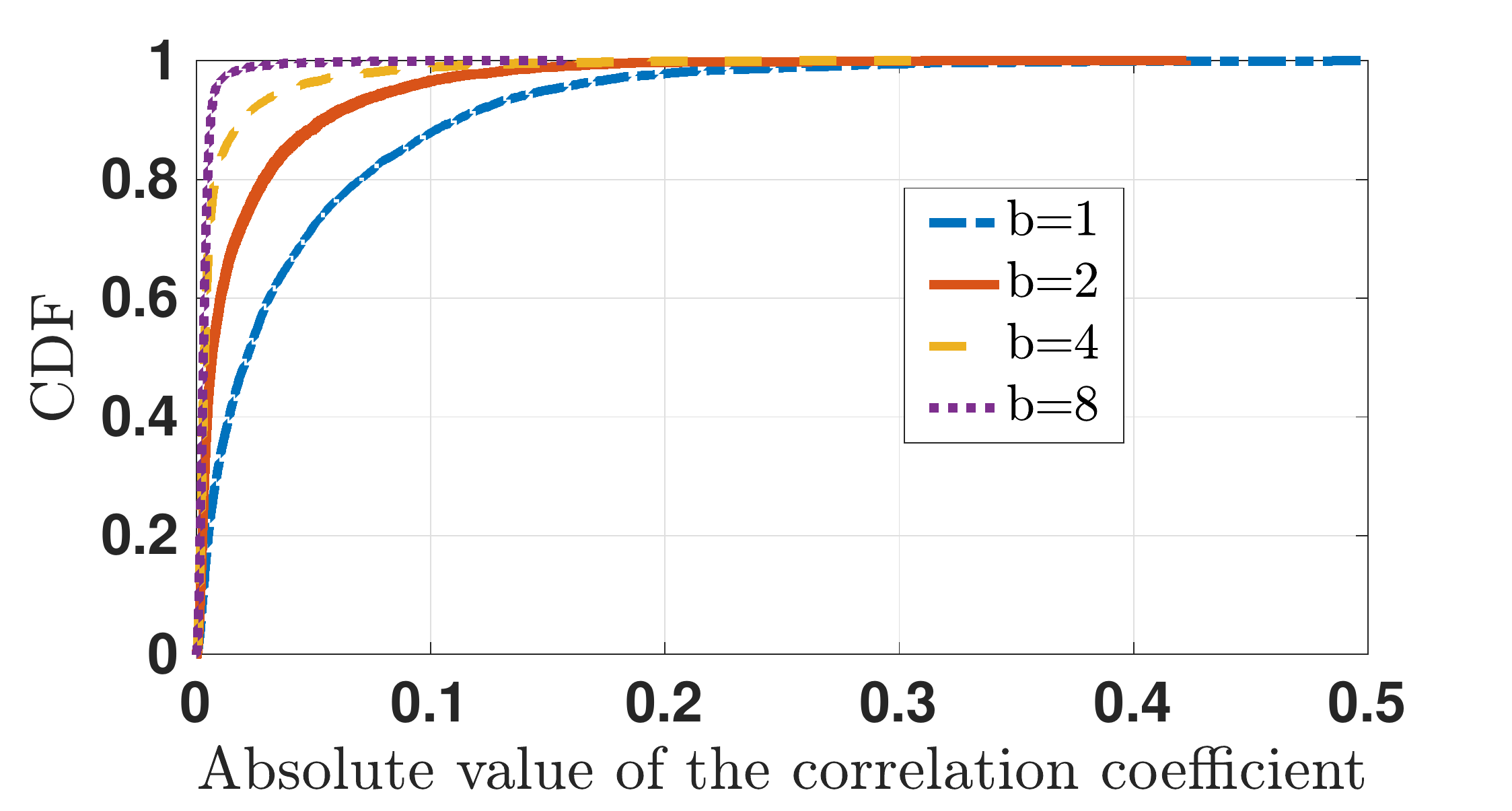}
		\vspace{-0.2cm}
		\caption{The CDF of the absolute value of the correlation coefficient between elements in $\bm{\eta}$.} \label{fig:sim3}
	\end{center}
	\vspace{-1.2cm}
\end{figure}

As an example, we consider a setup where a 4-antenna receiver quantifies the real and imaginary parts of each entry in the received signal $\vect{x}$ using identical $b$-bit ADCs. The input signal is generated as ${\bf x}={\bf H}{\bf s}$, where ${\bf H}\in \mathbb{C}^{4 \times 4}$ is the MIMO channel matrix from a 4-antenna transmitter.  We consider Rayleigh fading where  ${\bf H}$ has independent $\CN(0,1)$-distributed entries. For each channel realization, ${\bf H}$ is assumed perfectly known and the transmitted signal is ${\bf s}\sim \CN({\bf 0},{\bf I}_4)$, so ${\bf x}$ is conditionally complex Gaussian distributed. The Bussgang decomposition then says that the ADC output can be written as ${\bf z}={\bf D}{\bf x}+\bm{\eta}$. 
To demonstrate that the elements of $\bm{\eta}$ are correlated, Fig.~\ref{fig:sim3} shows the cumulative distribution function (CDF) of the normalized off-diagonal elements of ${\bf C}_{\eta}$ (i.e., the correlation coefficients) for different number of ADC bits.
When the ADC resolution is low, most of the correlation coefficients are non-zero and some are rather large. However, when the ADC resolution is high, the off-diagonal elements are almost zero and can potentially be approximated as zero when quantifying communication rates.

\vspace{-4mm}

\subsection{Generalized Bussgang Decomposition for Non-Gaussian Input Signals}
	\vspace{-0.1cm}
In the Bussgang theorem, we are utilizing that ${\bf x}$ and ${\bf y}$ are Gaussian signals. The main result cannot be generalized to non-Gaussian signals. However, we can always decompose the distorted signal according to \eqref{eq:Bussgang-decomp-MIMO} using the Bussgang gain matrix ${\bf B}={\bf C}_{zx}{\bf C}_{x}^{-1}$, but it generally won't be a diagonal matrix, even if an element-wise distortion of the type in \eqref{eq:elementwise-distortion} is used.
The intuition is that $\vect{B} \vect{x}$ is the linear MMSE estimate of $\vect{z}$ given a non-Gaussian distributed observation $\vect{x}$. In this analogy, $\bm{\eta}$ is the estimation error which is uncorrelated with $\vect{x}$ since
 \begin{align}
 \mathbb{E}\left\{\bm{\eta}{\bf x}^{\Htran}\right\}=\mathbb{E}\left\{\left({\bf z}-{\bf C}_{zx}{\bf C}_{x}^{-1}{\bf x}\right){\bf x}^{\Htran}\right\}=&{\bf C}_{zx}-{\bf C}_{zx}{\bf C}_{x}^{-1}{\bf C}_{x}={\bf 0}.
 \end{align}
The generalized Bussgang decomposition for non-Gaussian input ${\bf x}$ is illustrated in Fig.~\ref{fig:1}(b). It is suitable both for quantifying the SDR and to the analyze the performance of non-linear communication systems. For example, \cite{Demir2020} did this using practically modulated data signals. The paper also shows that although treating the uncorrelated distortion $\bm{\eta}$ as independent Gaussian noise is convenient, one can increase the performance by exploiting its information content.

\vspace{-3mm}

\section{Lessons Learned}

The Bussgang decomposition establishes that the output of a non-linear function is a scaled version of the random input signal plus an uncorrelated distortion term. It is an exact and unique representation. The distortion is not independent and not Gaussian, but can be treated as that to obtain a lower bound on the communication performance. The decomposition can be extended to MIMO systems but then the entries of the distortion vector are generally mutually correlated.

\vspace{-3mm}
\enlargethispage{3mm}

\bibliographystyle{IEEEtran}
\bibliography{IEEEabrv,refs}

\begin{thebibliography}{10}
\providecommand{\url}[1]{#1}
\csname url@samestyle\endcsname
\providecommand{\newblock}{\relax}
\providecommand{\bibinfo}[2]{#2}
\providecommand{\BIBentrySTDinterwordspacing}{\spaceskip=0pt\relax}
\providecommand{\BIBentryALTinterwordstretchfactor}{4}
\providecommand{\BIBentryALTinterwordspacing}{\spaceskip=\fontdimen2\font plus
\BIBentryALTinterwordstretchfactor\fontdimen3\font minus
  \fontdimen4\font\relax}
\providecommand{\BIBforeignlanguage}[2]{{%
\expandafter\ifx\csname l@#1\endcsname\relax
\typeout{** WARNING: IEEEtran.bst: No hyphenation pattern has been}%
\typeout{** loaded for the language `#1'. Using the pattern for}%
\typeout{** the default language instead.}%
\else
\language=\csname l@#1\endcsname
\fi
#2}}
\providecommand{\BIBdecl}{\relax}
\BIBdecl

\bibitem{Bussgang1952a}
J.~J. Bussgang, ``Crosscorrelation functions of amplitude-distorted {Gaussian}
  signals,'' Research Laboratory of Electronics, Massachusetts Institute of
  Technology, Tech. Rep. 216, 1952.

\bibitem{Minkoff1985a}
J.~{Minkoff}, ``The role of {AM-to-PM} conversion in memoryless nonlinear
  systems,'' \emph{{IEEE} Trans. Commun.}, vol.~33, no.~2, pp. 139--144, 1985.

\bibitem{Bjornson2019b}
E.~{Bj\"ornson}, L.~{Sanguinetti}, and J.~{Hoydis}, ``Hardware distortion
  correlation has negligible impact on {UL} massive {MIMO} spectral
  efficiency,'' \emph{{IEEE} Trans. Commun.}, vol.~67, no.~2, pp. 1085--1098,
  Feb 2019.

\bibitem{Schenk2008a}
T.~Schenk, \emph{{RF} imperfections in high-rate wireless systems: Impact and
  digital compensation}.\hskip 1em plus 0.5em minus 0.4em\relax Springer, 2008.

\bibitem{Fletcher2007a}
A.~K. Fletcher, S.~Rangan, V.~K. Goyal, and K.~Ramchandran, ``Robust predictive
  quantization: Analysis and design via convex optimization,'' \emph{IEEE
  Journal of Selected Topics in Signal Processing}, vol.~1, no.~4, pp.
  618--632, 2007.

\bibitem{Bai2013a}
Q.~Bai, A.~Mezghani, and J.~A. Nossek, ``On the optimization of {ADC}
  resolution in multi-antenna systems,'' in \emph{IEEE ISWCS}, 2013.

\bibitem{Hassibi2003a}
B.~Hassibi and B.~M. Hochwald, ``How much training is needed in
  multiple-antenna wireless links?'' \emph{{IEEE} Trans. Inform. Theory},
  vol.~49, no.~4, pp. 951--963, 2003.

\bibitem{papoulis2002}
A.~{Papoulis} and S.~U. {Pillai}, \emph{{Probability, Random Variables, and
  Stochastic Processes}}, 4th~ed.\hskip 1em plus 0.5em minus 0.4em\relax
  McGraw-Hill Higher Education, 2002.

\bibitem{McGee1969}
W.~{McGee}, ``Circularly complex {Gaussian} noise--a {Price} theorem and a
  {Mehler} expansion,'' \emph{IEEE Transactions on Information Theory},
  vol.~15, no.~2, pp. 317--319, 1969.

\bibitem{Demir2020}
{\"{O}}.~T. {Demir} and E.~{Bj\"{o}rnson}, ``Channel estimation in massive
  {MIMO} under hardware non-linearities: Bayesian methods versus deep
  learning,'' \emph{IEEE Open Journal of the Communications Society}, vol.~1,
  pp. 109--124, 2020.

\bibitem{Bjornson2014a}
E.~Bj{\"{o}}rnson, J.~Hoydis, M.~Kountouris, and M.~Debbah, ``Massive {MIMO}
  systems with non-ideal hardware: Energy efficiency, estimation, and capacity
  limits,'' \emph{{IEEE} Trans. Inform. Theory}, vol.~60, no.~11, pp.
  7112--7139, 2014.

\bibitem{Xu2019}
L.~{Xu}, X.~{Lu}, S.~{Jin}, F.~{Gao}, and Y.~{Zhu}, ``On the uplink achievable
  rate of massive {MIMO} system with low-resolution {ADC} and {RF}
  impairments,'' \emph{IEEE Communications Letters}, vol.~23, no.~3, pp.
  502--505, 2019.

\end{thebibliography}

\end{document}